# Recency effect disappears when information is integrated from independent perceptual sources

Sepide Bagheri[1], Mehdi Keramati[2], Reza Ebrahimpour[3,4], Sajjad Zabbah[3,5,6*]

[1]Department of Psychology, University of Tehran, Tehran, Iran

[2]Department of Psychology and Neuroscience, School of Health and Medical Sciences, City St George's University of London, London, UK

[3]Center for Cognitive Science, Institute for Convergence Science and Technology (ICST), Sharif University of Technology, Tehran, Iran

[4]School of Cognitive Sciences, Institute for Research in Fundamental Sciences (IPM), Tehran, Iran

[5]Wellcome Centre for Human Neuroimaging, UCL Queen Square Institute of Neurology, University College London, 12 Queen Square, London WC1N 3AR, United Kingdom

[6]Department of Computing, Goldsmiths, University of London, SE14 6NW London, United Kingdom

*Corresponding author email: s.zabbah@ucl.ac.uk

# Abstract

Decision-making often involves integrating discrete pieces of information from distinct sources over time, yet the cognitive mechanisms underlying this integration remain unclear. In this study, we examined how individuals accumulate and integrate discrete sensory evidence. Participants performed a task involving random dot motion stimuli presented in single- and double-pulse trials. These stimuli varied in motion coherence, source consistency of pulses (either the same or orthogonal directions), and temporal gaps between pulses. We found that participants effectively integrated information regardless of source type or temporal gaps. As expected, when both pulses originated from the same source, performance showed a sequence-dependent effect—accuracy was influenced by the order of pulse presentation. However, this effect disappeared when the pulses came from orthogonal sources. Confidence judgments were similarly unaffected by temporal gaps or pulse sequence but were higher when information originated from orthogonal sources. These findings highlight the specific role of perceptual independency on information integration and consequently, on decision-making.

# Introduction

Decision-making is the process of selecting a choice among alternatives, which underscores how much of our daily behavior arises from decision-making processes. To make a decision, studies suggest that the brain accumulates evidence over time until reaching a threshold, known as the "decision bound" (Gold & Shadlen, 2007; Ratcliff & McKoon, 2008). While most studies have focused on the continuous stream of evidence from a single source, real-life decisions often rely on multiple distinct sources of information that are often separated by temporal gaps. However, the mechanisms governing decision-making when integrating discrete and distinct sources of evidence remain less understood.

Several studies have investigated how decision-making operates when discrete pieces of evidence originating from a single source are separated by temporal gaps. These studies have shown that human participants can optimally accumulate discrete evidence to improve their decisions, regardless of whether the temporal gaps between them are short (Kiani et al., 2013) or long (Waskom & Kiani, 2018). However, even when participants are well-trained on continuous stimuli (e.g., random dot motion), a sequence-dependent effect has been observed, meaning that the order in which evidence is received influences behavior (Kiani et al., 2013; Tohidi-Moghaddam et al., 2019). This sequence-dependency has been attributed to a time-dependent rate of evidence accumulation (i.e., time-dependent attentional signals or drift rate) (Tohidi-Moghaddam et al., 2019). Neural substrates of the accumulation of discrete evidence has been found in the lateral intraparietal (LIP) cortex (Kira et al., 2015). While some studies have examined how perceptual accuracy changes in discrete environments, fewer have explored how confidence is affected. Recent work by Azizi and Ebrahimpour (2023) showed that confidence, like accuracy, remains unaffected by the temporal gap between evidence pulses.

In this study, we aim to investigate the underlying mechanisms of decision-making when discrete evidence comes from distinct sources, as is common in many real-life situations. To this end, we designed a perceptual decision-making task using random dot motion (RDM) stimuli. The experiment consists of single- and double-pulse trials, where the pulses originate from either similar or different types of sources. Participants were asked to report both the direction of motion in the stimulus as well as their level of confidence in their decision.

Our findings reveal that (1) participants can improve their decisions using information from distinct sources, regardless of the temporal gap between them; (2) the sequence-dependency effect is influenced by the type of information source; (3) participants perform more accurately than predicted by a perfect accumulator model; and (4) the source-type effect is also observed in participants' confidence, and just like accuracy, confidence is unaffected by the temporal gap.

# Materials and Methods

## Participants

The behavioral experiment included four participants (two males and two females) aged between 24 and 34. All participants had normal or corrected-to-normal vision and were unaware of the experiment's purpose. Informed consent was obtained from all participants prior to their involvement. The study was approved by the Ethics Committee of Iran University of Medical Sciences (IR.IUMS.REC.1399.1282.).

## Experimental task

The task comprised 8 blocks (12 blocks for Subject 4), each containing 150 trials. Participants could take sufficient breaks between blocks (see Supplementary Materials for details). In each trial, they observed a collection of dots presented in either one (single-pulse) or two (double-pulse) consecutive short intervals (600ms each) and decided whether the dots were moving toward the right (R)/up (U) or the left (L)/down (D) (Fig. 1).

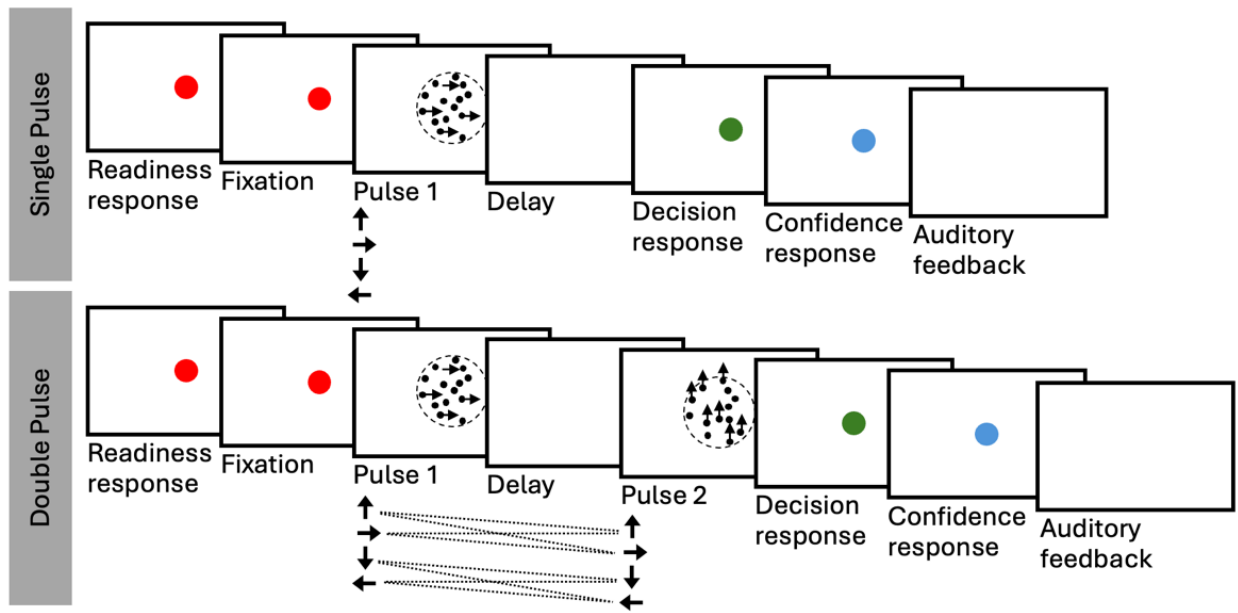

**Figure 1.** Direction discrimination task with single- and double-pulse random dot motions.

In a single-pulse trial, the participant indicated readiness by pressing the spacebar on the keyboard after a red fixation point appeared at the center of the screen. Following this readiness response, a fixation gap of 200–500ms occurred, after which the red fixation point was replaced by a random dot stimulus displayed for 600ms (See Supplementary Methods for details). The motion coherence of the stimulus was randomly selected from 0%, 3.2%, 6.4%, 12.8%, or 51.2%, and the motion direction was randomly assigned to one of four options: Up, Right, Down, or Left. Once the motion ended, the screen turned black for either 500ms (gap 0.5) or 1000ms (gap 1). After this delay, a green dot appeared at the center of the screen, prompting the participant to press "D" if the motion direction was U or R, or "Z" if the direction was D or L. Upon making this decision, the green dot was replaced by a blue dot, signaling the participant to indicate their confidence level by pressing "G" for high confidence or "B" for low confidence. Auditory feedback was provided to indicate whether the response was correct or incorrect. In total, a single-pulse trial could be one of 48 types, determined by 6 (coherence levels) × 4 (directions) × 2 (delay durations).

In a double-pulse trial, the motion coherence of the first pulse was randomly selected from three values: 0%, 6.4%, or 12.8%. After the first pulse, a delay of either 500ms or 1000ms followed, during which the screen turned black. This was then followed by a second pulse, whose coherence level was randomly selected from two values that were not used for the first pulse. In other words, the coherence of the second pulse was always either weaker or stronger than the first pulse, but never the same. The motion direction of the second pulse depended on the first pulse: if the first pulse was U or R, the second pulse was randomly assigned as either U or R; similarly, if the first pulse was D or L, the second pulse was randomly assigned as either D or L. In other words, the motion directions of the two pulses were always consistent: they were both selected from either {R, U} or {L, D} sets. Thus, participants could accumulate and integrate information across the two pulses, even when the pulses had orthogonal motion directions. We label this as "different" source types, as opposed to "same" source type where the second pulse has the same direction as the first. As in single-pulse trials, participants pressed "D" or "Z" in response to R/U or D/L directions, respectively, upon seeing the green dot. Upon seeing the blue dot, they reported their confidence level by pressing "G" or "B" for high or low confidence, respectively, and then received auditory feedback. A double-pulse trial could be one of 96 types, determined by 3 (first pulse's coherence) × 4 (first pulse's directions) × 2 (delay durations) × 2 (second pulse's coherence) × 2 (second pulse's directions).

The two trial types created a total of 144 (48 + 96) conditions. These were randomly arranged in 1200 trials (1800 for Subject 4), with each condition being presented at least four times. Prior to that, participants underwent extensive pre-training to familiarize themselves with the task. They were instructed to respond as accurately and quickly as possible to the presented stimuli.

## Data analysis

A variety of logistic regression models were used to assess the influence of stimulus parameters on binary outcomes. The effect of motion coherence, $C$, on the probability of a correct choice, $P_{correct}$, and the probability of expressing high confidence, $P_{confidence}$, was modeled as follows for single-pulse trials:

$$Logit\,(P_{correct}) = \beta_0 + \beta_1 C + \varepsilon \qquad (1)$$

$$Logit(P_{confidence}) = \beta_0 + \beta_1 C + \varepsilon \qquad (2)$$

where $Logit\,[P]$ is defined by $log\left(\frac{P}{1-P}\right)$, and $\varepsilon$ is the error term.

To evaluate potential bias in performance, a similar equation was used

$$Logit\ (P_{options}) = \beta_0 + \beta_1 C + \varepsilon \qquad (3)$$

where motion in upward or rightward directions indicated a choice for option one (reported by pressing the "D" key), while motion in leftward or downward directions indicated a choice for option two (reported by pressing the "Z" key). The null hypothesis was that no bias exists.

To examine the effect of the inter-pulse interval on performance during double-pulse trials, a more complex model was used:

$$Logit\ (P_{correct}) = \beta_0 + \beta_1 C_1 + \beta_2 C_2 + \beta_3 T + \beta_4 D + \beta_5 C_1 T + \beta_6 C_2 T + \beta_7 TD + \varepsilon \qquad (4)$$

where $C_1$ and $C_2$ are the motion coherences of the first and second pulses, and $T$ is the inter-pulse interval. $D$ represents the source type (same or different), indicating whether the motion direction of the second pulse was the "same" as the first pulse, or "different". The null hypothesis was that neither the temporal gap nor the source type significantly affects accuracy.

A similar model was used to evaluate confidence:

$$Logit(P_{confidence}) = \beta_0 + \beta_1 C_1 + \beta_2 C_2 + \beta_3 T + \beta_4 D + \beta_5 C_1 T + \beta_6 C_2 T + \beta_7 TD + \varepsilon \qquad (5)$$

with the null hypothesis that temporal gap and source type do not influence confidence.

To investigate the impact of pulse sequence and source type on accuracy, the following model was used:

$$Logit\ (P_{correct}) = \beta_0 + \beta_1 S + \beta_2 D + \beta_3 SD + \varepsilon \qquad (6)$$

where $S$, representing pulse sequence, indicates whether the first pulse was stronger than the second pulse (strong-weak) or the opposite (weak-strong). $D$ represent the source type. The null hypothesis was that accuracy is independent of both pulse sequence and source type. The same model was applied to confidence:

$$Logit(P_{confidence}) = \beta_0 + \beta_1 S + \beta_2 D + \beta_3 SD + \varepsilon \qquad (7)$$

with the null hypothesis that confidence is not affected by pulse sequence or source type.

To estimate expected accuracy in double-pulse trials under the assumption of optimal integration, we derived an estimate based on the evidence provided by each pulse. This estimate assumes that the decision is based on the sum of evidence from the two pulses, with the evidence distribution inferred from the accuracy of single-pulse trials. For each pulse, the evidence ($e_1$ and $e_2$) was inferred from the corresponding accuracy ($P_1$ and $P_2$; where $P$ was calculated per participant and per motion coherence level) using the inverse of the cumulative normal distribution function:

$$e_1 = \phi^{-1}(P_1|\mu = 0, \sigma = 1) \quad , \quad e_2 = \phi^{-1}(P_2|\mu = 0, \sigma = 1),$$

$$\phi(s|\mu, \sigma) = \int_{-\infty}^{s} N(v|\mu, \sigma) dv,$$

where $\phi$ represents the cumulative normal distribution function, and $\phi^{-1}$ is its inverse. The expected accuracy for double-pulse trials was then computed as:

$$P_e = 1 - \phi(0|\mu = e_1 + e_2, \sigma = \sqrt{2}).$$

This expected accuracy ($P_e$) was compared to the observed accuracy in double-pulse trials using logistic regression:

$$Logit\ (P_{correct}) = Logit(P_e) + \beta + \varepsilon \tag{8}$$

A significant positive $\beta$ would indicate that the actual accuracy exceeds the expected accuracy.

## Results

Trials containing single and double pulses were randomly intermixed throughout the experiment. Similarly, the characteristics of each pulse, such as its strength, direction, and inter-pulse interval varied randomly across trials. Before the main experiment, participants underwent extensive training in all motion directions to ensure they reached a consistent and high level of accuracy in their responses (See Supplementary Methods for details). During the last training session, only subject 3 showed significant choice bias (equation (3), $\beta_0 = -0.231, p = 0.0015$; Table S1) while no significant choice bias was found in any participant's choice accuracy during the main experiment (equation (3), $\beta_0 = 0.092, p > 0.05$, Table S2). As expected, participants were more accurate under stronger signals (Fig. 2**a**; equation (1), $\beta_1 > 0, p < 0.0001$, Table 3; equation (4), $\beta_1 = 5.313, p < 0.0001, \beta_2 = 4.629, p < 0.0001$, Table 5).

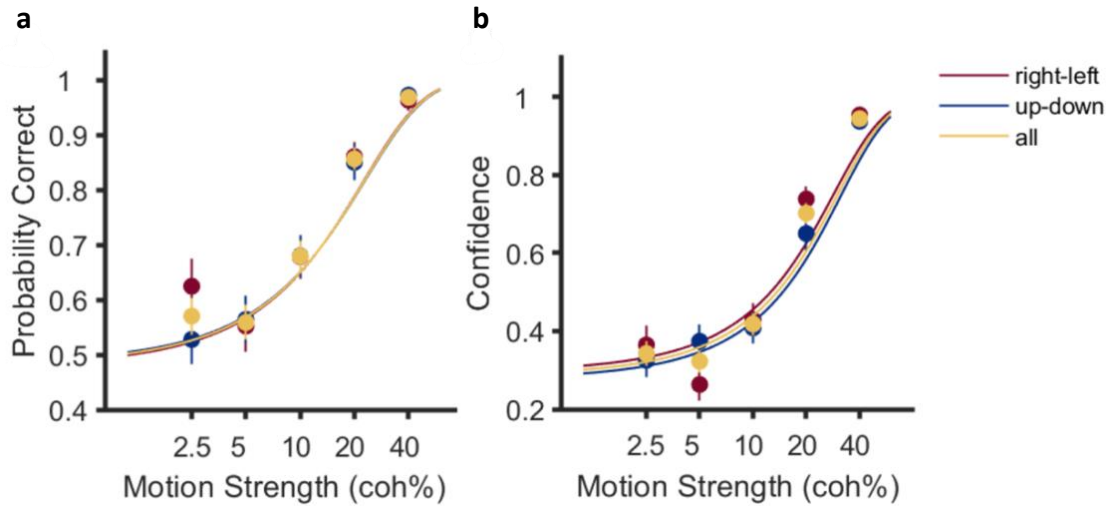

**Figure 2.** Accuracy and confidence of responses as a function of stimulus strength.

At the end of each trial, participants were asked to report their confidence in the likelihood that their response was correct. In line with their choice accuracy, participants were more confident in their responses under high motion strength conditions (Fig. 2**b**; equation (2), $\beta_1 > 0, p < 0.0001$, Table 4).

## Accuracy is independent of source type and the temporal gap between pulses

Consistent with previous research (Kiani, 2013; Azizi, 2021), our results show that the temporal gap between pulses of information did not significantly affect participants' accuracy (equation (4), $\beta_3 = 0.184, p = 0.278$; Table S5). More importantly, this gap-independent behavior persisted regardless of the information source type (i.e., whether the directions of the two pulses were the same or orthogonal). In Equation 4, no interaction between the inter-pulse interval and the source type was observed ($\beta_7 = -0.005, p = 0.97$; Table S5), suggesting that participants integrated discrete pulses of information from different sources, irrespective of the temporal gap between them. In other words, this gap-independent behavior occurred even when information came from different sources (Fig. 3**a**).

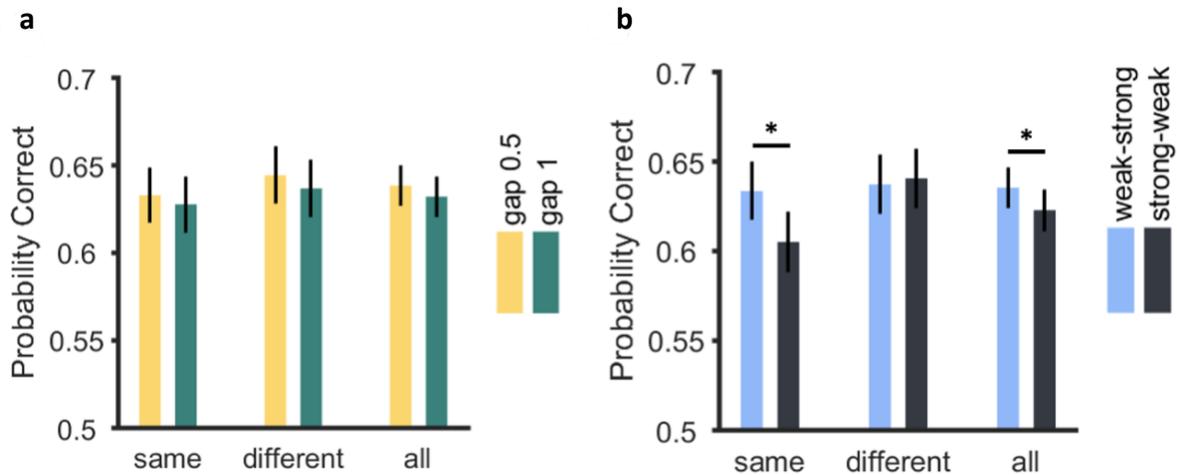

**Figure 3.** Gap-independent and sequence-dependent effects of source types on choice accuracy.

## Sequence-dependent behavior is influenced by source type

Previous research has shown that decision accuracy with temporally discrete cues is influenced by the sequence in which those cues are presented, with stronger accuracy observed when the final pulse is the stronger one (Kiani et al., 2013; Tohidi-Moghaddam et al., 2019; Azizi & Ebrahimpour, 2023). Our results indicate that this effect is modulated by the consistency of the information source. When both pulses come from the same source (i.e., both motions have the same direction), the sequence-dependent accuracy shows a recency effect (Fig. 3**b**; Table S7), consistent with previously reported findings (Kiani et al., 2013; Tohidi-Moghaddam et al., 2019; Azizi & Ebrahimpour, 2023). However, when participants combine information from different sources, this sequence-dependent behavior disappears (Fig. 3**b**; Table S7); this pattern holds for different combinations of coherence levels (0%-6.4%, 0%-12.8%, and 6.4%-12.8%; see Fig. S1). Furthermore, source type has a significant effect on the extent to which weak-strong vs. strong-weak sequences affect accuracy (equation (6), $\beta_3 = 0.335$, $p = 0.017$; Table S7). This suggests that the putative recency effect does not readily generalize to conditions where information comes from source types with different sensory-level mechanisms.

## Regardless of source type, participants' accuracy is higher than predicted by perfect accumulator model

An effect of better-than-expected accuracy has been observed when discrete cues originate from a single information source (Majidpour et al., 2025; Tohidi-Moghaddam et al., 2019; Kiani et al., 2013). Using participants' performance in the single-pulse trials, we predicted their accuracy in double-pulse trials based on the assumption of the perfect integrator model. As observed in previous studies, participants outperformed the predictions of the perfect accumulator model. This effect was consistent regardless of whether information came from the same or different sources (Fig 4; equation (8), $\beta = 0.209$, $p < 0.0001$; Tables S9-S11).

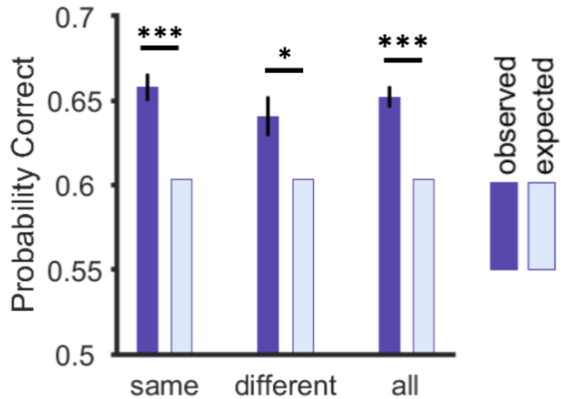

**Figure 4.** Comparing participants' accuracy in double-pulse trials with that of a perfect accumulator model.

### Confidence is dependent on source type but independent of temporal gap

Like accuracy, reported confidence was not influenced by the temporal gap between pulses (Fig. 5**a**; equation (5), $\beta_3 = -0.29, p = 0.07$; Table S6). This temporal independence of confidence aligns with the effect reported previously by Azizi and Ebrahimpour (2023). Moreover, confidence was not affected by the sequence-type of pulses. In other words, there was no significant effect on reported confidence between the strong-weak and weak-strong conditions (Fig. 5**b**; equation (7), $\beta_1 = 0.1, p = 0.3$; Table S8). However, in contrast to our prediction, confidence was higher when evidence was gathered from different sources (i.e., orthogonal directions) than when the two pulses had the same source type (i.e., same direction) (equation (7), $\beta_2 = 0.23, p = 0.017$; Table S8). These findings highlight distinct mechanisms for confidence formation compared to choice selection.

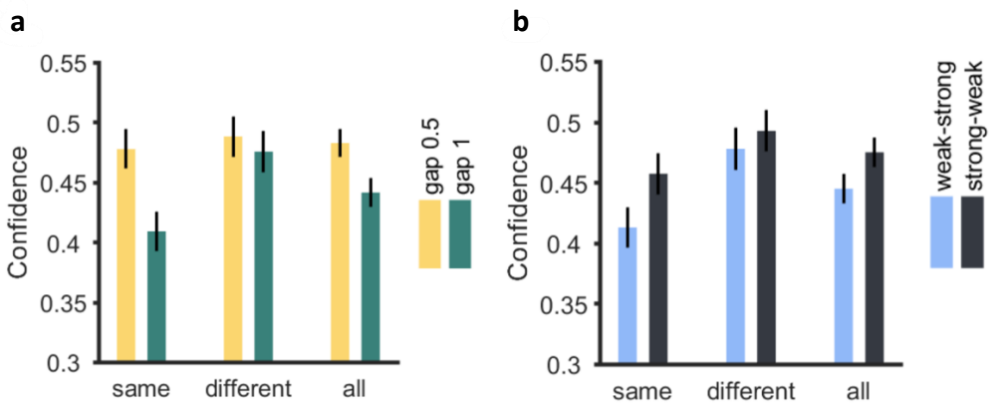

**Figure 5.** Effect of temporal gap and sequence-type on reported confidence.

# Discussion

In this study, we investigated how decision-making operates when discrete pieces of evidence come from distinct sources and are separated by temporal gaps. Our findings, in line with previous literature (Azizi et al., 2021; Azizi & Ebrahimpour, 2023), demonstrate that participants can effectively integrate information from different sources, maintain stable accuracy and confidence across varying temporal gaps. Although the accuracy and confidence remain intact over different temporal gaps, they exhibit a sequence-dependent effect when discrete pieces of evidence come from same sources. Specifically, participants perform better when the strong pulse appears at the end, compared to when it appears at the beginning. The sequence-dependent effect, however, vanishes if evidence is accumulated from distinct sources. Moreover, the better-than-expected-accuracy effect (i.e., participants performed more accurately than predicted by a perfect accumulator model), which has been observed in previous studies with same sources (Kiani et al., 2013; Tohidi-Moghaddam et al., 2019), still holds when information comes from different sources.

One key finding in our study is that accuracy remained independent of the temporal gap between pulses, regardless of whether the two pulses originated from the same or different sources. This is consistent with previous studies showing that human participants can optimally accumulate discrete evidence over time to improve their decisions, regardless of whether the gap between them is short or long (Kiani et al., 2013; Waskom et al., 2018). Our results extend these findings by demonstrating that even when evidence comes from different sources, participants can integrate this information without temporal interference, suggesting a robust mechanism for discrete evidence accumulation. Our results imply that such integration takes place at neural levels beyond those specific to individual sources. The independence from temporal gap also aligns with prior findings that confidence judgments remain stable regardless of the interval between evidence pulses (Azizi & Ebrahimpour, 2023). These results suggest that decision-making mechanisms are well adapted to situations where information is presented discontinuously. Additionally, our findings are in line with Majidpour et al. (2025), who observed similar gap-independent behavior when evidence was presented at the center of visual field.

While accuracy remained stable across temporal gaps, we observed that the sequence dependent effect was influenced by the consistency of the information source. In line with prior work (Kiani et al., 2013; Tohidi-Moghaddam et al., 2019; Azizi et al., 2023), accuracy was higher when a strong pulse followed a weaker one; a phenomenon known as recency effect (Rahnev & Denison, 2018; Cheadle et al., 2014). However, this effect was reported only when both pulses came from the same source. As a potential underlying mechanism for such recency effect, Wyart et al. (2012b) suggested that observers should exhibit higher sensitivity to pieces of evidence that are in line with the priors; a possible mechanism that is supported by Cheadle et al. (2014) on human participants in a discrete perceptual decision-making task while the source of evidence remains constant in the course of each trial. Cheadle et al. (2014) proposed an adaptive gain mechanism, which posits that later-occurring evidence is processed with higher gain compared to earlier evidence, leading to a recency bias. This mechanism explains why we observed stronger influence from the second pulse when both pulses originated from the same source, as belief-consistent information may be given greater weight. The adaptive gain model suggests that this process is not merely a consequence of memory decay but an active modulation of information processing. This may also explain why the recency effect disappears when pulses originate from different sources, as conflicting or less congruent information may not receive the same level of gain modulation. In any case, the fact that recency effect vanishes in heterogenous trials means that adaptive gain control model is sensitive to the source of information.

While our study shows the dependency of adaptive gain on source consistency, it is also demonstrated to require consistent spatial location of the information source. Majidpour et al. (2025) showed that recency effect disappears if discrete pulses in random dot motion tasks appear in different spatial location.

Another critical finding of our study is that participants outperformed the predictions of a perfect accumulator model, regardless of whether the information came from the same or different sources. This effect has been previously reported in cases where discrete cues originate from a single information source type (Kiani et al., 2013; Tohidi-Moghaddam et al., 2019). Our findings indicate that this better-than-expected performance generalizes to conditions where information is gathered from distinct sources. This suggests that beyond simple evidence accumulation, additional cognitive processes—such as enhanced attentional engagement, strategic decision weighting, or dynamic accumulation adjustments—may contribute to improved accuracy.

Finally, our results indicate that confidence judgments are influenced by the type of information source, but remain independent of both the temporal gap and the order in which pulses are presented. Unlike previous work showing that accuracy follows a sequence-dependent pattern when evidence originates from a single source (Kiani et al., 2013; Tohidi-Moghaddam et al., 2019), we did not observe a similar effect for confidence. Instead, confidence was higher when the two pulses came from different sources, suggesting a distinct mechanism for confidence formation compared to choice selection. This may indicate that participants perceive evidence from different sources as providing complementary or reinforcing information, leading to an overall increase in subjective certainty. Future research should explore whether confidence relies on a separate accumulation process or if metacognitive judgments incorporate additional cues beyond those used for choice accuracy.

# Data Availability

https://github.com/sepidbagheri/Decision-making-with-Discrete-Sources/tree/main/Data

# Code Availability

https://github.com/sepidbagheri/Decision-making-with-Discrete-Sources/tree/main/Code

# Competing interests

The authors declare no competing interests.

# Figure Legends

**Figure 1.** Direction discrimination task with single- and double-pulse random dot motions. In single-pulse trials, the motion was randomly chosen among four direction conditions (U, D, L, R) and six coherence levels (0%, 6.4%, 12.8%, 25.6%, 51.2%), and the delay was chosen from two conditions (500 or 1000 msec). In double-pulse trials, the first motion pulse was randomly chosen among four direction conditions (U, D, L, R), and three coherence levels (0%, 6.4%, 12.8%), and the delay was chosen from two conditions (500 or 1000 msec); the second pulse was randomly selected from U and R if the first pulse was U or R, and from D or L if the first pulse was D or L. The coherence of the second pulse was always different from that of the first pulse; either weaker or stronger. In both trial types, participants were instructed to press "D" in repose to U or R directions, and "Z" in repose to D or L directions.

**Figure 2.** Accuracy and confidence of responses as a function of stimulus strength. (**a**) Probability of correct responses in single-pulse trials for stimuli moving in the right/left (red) and up/down (blue) directions. The yellow curve represents averaged data across all directions. (**b**) Average confidence in responses in single-pulse trials. Error bars represent standard error.

**Figure 3.** Gap-independent and sequence-dependent effects of source types on choice accuracy. (**a**) Participants' accuracy in double-pulse trials was independent of the temporal gap between pulses, regardless of the type of information source. (**b**) Sequence-dependent behavior varied based on the information source type: in the same source condition, accuracy improves when a strong pulse follows a weaker one; when the information comes from different sources, however, the sequence-dependent effect disappears.

**Figure 4.** Comparing participants' accuracy in double-pulse trials with that of a perfect accumulator model. Participants consistently make more accurate choices than those predicted by the perfect accumulator model, regardless of the information source.

**Figure 5.** Effect of temporal gap and sequence-type on reported confidence. (**a**) Reported confidence was not affected by the temporal gap between pulses, regardless of whether the information came from the same or different sources. (**b**) No sequence dependency effect was observed in reported confidence. However, the reported confidence was higher when the sources were different compared to when they were the same.

# Supplementary Methods

## Stimulus

The stimulus consisted of dynamic random dots with a density of 16.7 dots/degree²/s. Each dot was represented by a 2x2 white square pixel, moving in one of four directions depending on the stimulus coherence.

## Pretraining

Participants attended in multiple training sessions until they reached an accuracy of approximately 80–86% in their responses. Each participant took part in up to three sessions per day. Each session consisted of two blocks, and each block contained 150 trial; this mounted to a total of 300 trials per session. In this phase of the experiment, the aim was to help participants adjust their decision-making criterion to the appropriate state.

In the training phase, all trials within one block came from the same category in an alternating manner. At the start of each block, participants were informed whether the upcoming stimuli would involve up–down or left–right motion. For each participant, the order of the blocks was randomly selected. For instance, a participant might have first completed a block with left–right stimuli and then a block with up–down stimuli, or vice versa.

## Main experiment

In the main part of the experiment, participants performed 1,200 trials (1,800 trials for Participant 4), which were split into 4 blocks in total; except for Participant 4—who completed 6 sessions—each participant completed 4 sessions, with each session comprising 2 blocks. Each block contained 150 trials, making a total of 300 trials per session. Participants took approximately a 10-minute break between blocks to avoid fatigue.

# Supplementary material

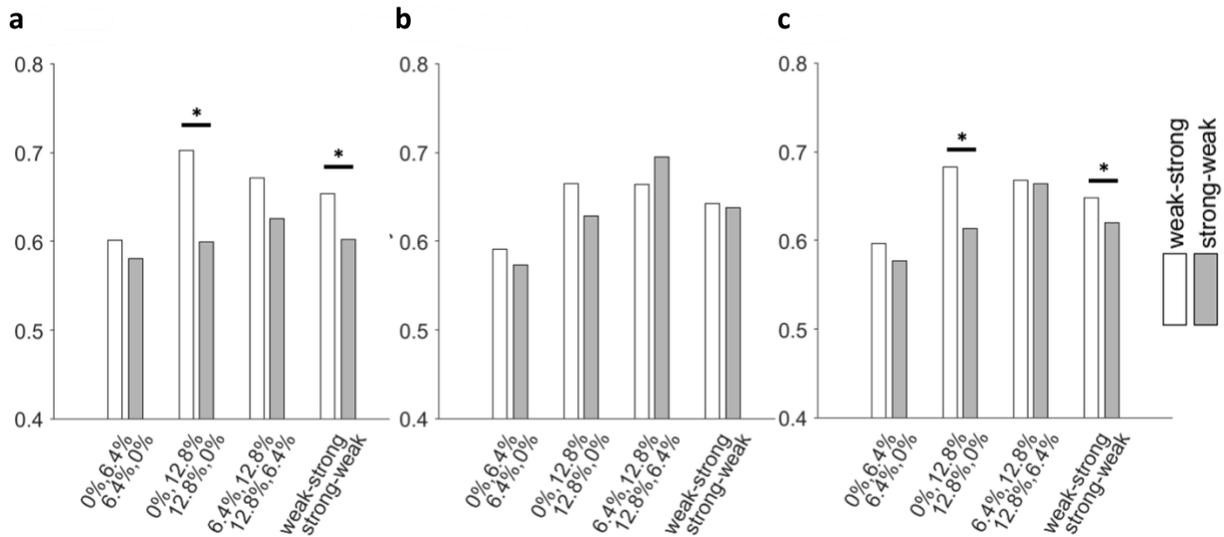

**Figure S1.** Sequence-dependent behavior varied based on the information source type. (**a**) In the same source condition, accuracy improves when the weakest pulse type (0%) is followed by the strongest pulse type (12.8%). This pattern stays present in the pooled condition (i.e., the fourth pair of bars). (**b**) When the information comes from different sources, however, the sequence-dependent effect disappears. (**c**) The pattern in the same-type condition is pronounced enough to stay present when the two conditions (same and different) are pooled together.

**Table S1.** Choice bias in train part, based on equation (3)

| Subjects | $\beta_0$ |
| --- | --- |
| Sub 1 | -0.144 ($p > 0.05$) |
| Sub 2 | -0.280 ($p > 0.05$) |
| Sub 3 | -0.231 ($p < 0.01$) |
| Sub 4 | -0.046 ($p > 0.05$) |
| pool | -0.173 ($p = 0.0015$) |

**Table S2.** Choice bias in main part, based on equation (3)

| Subjects | $\beta_0$ |
| --- | --- |

| Sub 1 | 0.191 ($p > 0.05$) |
| Sub 2 | 0.241 ($p > 0.05$) |
| Sub 3 | -0.100 ($p > 0.05$) |
| Sub 4 | -0.027 ($p > 0.05$) |
| pool | 0.092 ($p > 0.05$) |

**Table S3.** Accuracy and Coherency, based on equation (1)

| Subjects | $\beta_1$ |
|---|---|
| Sub 1 | 6.853 ($p < 0.0001$) |
| Sub 2 | 8.012 ($p < 0.0001$) |
| Sub 3 | 10.873 ($p < 0.0001$) |
| Sub 4 | 5.680 ($p < 0.0001$) |
| pool | 6.853 ($p < 0.0001$) |

**Table S4.** Confidence and Coherency, based on equation (2)

| Subjects | $\beta_1$ |
|---|---|
| Sub 1 | 5.058 ($p < 0.0001$) |
| Sub 2 | 8.858 ($p < 0.0001$) |
| Sub 3 | 8.097 ($p < 0.0001$) |
| Sub 4 | 6.907 ($p < 0.0001$) |
| pool | 6.708 ($p < 0.0001$) |

**Table S5.** probability of correct, based on equation (4)

| Subjects | $\beta_1$ | $\beta_2$ | $\beta_3$ | $\beta_4$ | $\beta_5$ | $\beta_6$ | $\beta_7$ |
|---|---|---|---|---|---|---|---|
| Sub 1 | 3.263 | 3.833 | 0.349 | 0.081 | −7.499 | 1.070 | 0.163 |

|       | (p = 0.169)   | (p = 0.139)   | (p = 0.407) | (p = 0.715) | (p = 0.026) | (p = 0.767) | (p = 0.607) |
|-------|---------------|---------------|-------------|-------------|-------------|-------------|-------------|
| Sub 2 | 5.664         | 9.962         | 0.738       | 0.364       | −5.796      | −5.989      | 0.021       |
|       | (p = 0.019)   | (p < 0.0001)  | (p = 0.082) | (p = 0.100) | (p = 0.083) | (p = 0.090) | (p = 0.943) |
| Sub 3 | 7.815         | 5.742         | −0.016      | 0.013       | −1.390      | 2.010       | −0.007      |
|       | (p < 0.0001)  | (p = 0.003)   | (p = 0.952) | (p = 0.949) | (p = 0.540) | (p = 0.481) | (p = 0.979) |
| Sub 4 | 2.372         | 0.310         | 0.111       | −0.104      | −2.411      | 0.533       | −0.118      |
|       | (p = 0.225)   | (p = 0.871)   | (p = 0.747) | (p = 0.549) | (p = 0.360) | (p = 0.842) | (p = 0.624) |
| pool  | 5.313         | 4.629         | 0.184       | 0.027       | −3.122      | −0.123      | −0.005      |
|       | (p < 0.0001)  | (p < 0.0001)  | (p = 0.278) | (p = 0.78)  | (p = 0.011) | (p = 0.933) | (p = 0.971) |

**Table S6.** Confidence, based on equation (5)

| Subjects | $\beta_1$ | $\beta_2$ | $\beta_3$ | $\beta_4$ | $\beta_5$ | $\beta_6$ | $\beta_7$ |
|----------|-----------|-----------|-----------|-----------|-----------|-----------|-----------|
| Sub 1    | −0.724    | 2.460     | −0.515    | 0.093     | 5.875     | −0.379    | 0.120     |
|          | (p = 0.798) | (p = 0.427) | (p = 0.290) | (p = 0.726) | (p = 0.138) | (p = 0.928) | (p = 0.747) |
| Sub 2    | 6.094     | 5.793     | −0.339    | −0.149    | −0.594    | −0.257    | 0.308     |
|          | (p = 0.007) | (p = 0.014) | (p = 0.416) | (p = 0.474) | (p = 0.852) | (p = 0.938) | (p = 0.295) |
| Sub 3    | 5.984     | 6.194     | −0.422    | −0.214    | 0.694     | −0.785    | 0.388     |
|          | (p < 0.0001) | (p = 0.0009) | (p = 0.155) | (p = 0.290) | (p = 0.691) | (p = 0.775) | (p = 0.183) |
| Sub 4    | 3.744     | 3.024     | −0.299    | 0.234     | 1.675     | −0.697    | 0.219     |
|          | (p = 0.384) | (p = 0.803) | (p = 0.543) | (p = 0.190) | (p = 0.422) | (p = 0.124) | (p = 0.062) |
| pool     | 4.465     | 4.443     | −0.299    | 0.029     | 0.827     | −0.877    | 0.253     |
|          | (p < 0.0001) | (p < 0.0001) | (p = 0.079) | (p = 0.755) | (p = 0.481) | (p = 0.541) | (p = 0.063) |

**Table S7.** Sequence dependency in the probability of correct, based on equation (6)

| Subjects | $\beta_1$ | $\beta_2$ | $\beta_3$ |
|----------|-----------|-----------|-----------|
| Sub 1    | −0.434 (p = 0.051) | −0.061 (p = 0.784) | 0.448 (p = 0.149) |
| Sub 2    | −0.317 (p = 0.134) | 0.453 (p = 0.040)  | −0.061 (p = 0.842) |
| Sub 3    | −0.462 (p = 0.038) | −0.157 (p = 0.512) | 0.517 (p = 0.081) |
| Sub 4    | −0.118 (p = 0.495) | −0.318 (p = 0.061) | 0.328 (p = 0.173) |
| pool | −0.0.287 (p = 0.004) | −0.050 (p = 0.622) | 0.335 (p = 0.016) |

**Table S8.** Sequence dependency in confidence, based on equation (7)

| Subjects | $\beta_1$ | $\beta_2$ | $\beta_3$ |
|---|---|---|---|
| Sub 1 | 0.106 (p = 0.677) | 0.240 (p = 0.351) | −0.073 (p = 0.841) |
| Sub 2 | 0.044 (p = 0.830) | −0.011 (p = 0.954) | 0.109 (p = 0.706) |
| Sub 3 | 0.313 (p = 0.157) | 0.538 (p = 0.020) | −0.521 (p = 0.070) |
| Sub 4 | 0.077 (p = 0.673) | 0.263 (p = 0.139) | 0.157 (p = 0.529) |
| pool | 0.101 (p = 0.304) | 0.234 (p = 0.018) | −0.135 (p = 0.317) |

**Table S9.** Observed vs. expected correctness based on source type

| | same source | | different source | | all | |
|---|---|---|---|---|---|---|
| Subjects | observed correctness | expected correctness | observed correctness | expected correctness | observed correctness | expected correctness |
| Sub 1 | 0.672 | 0.677 | 0.638 | 0.675 | 0.661 | 0.676 |
| Sub 2 | 0.645 | 0.551 | 0.705 | 0.551 | 0.665 | 0.551 |
| Sub 3 | 0.690 | 0.673 | 0.674 | 0.658 | 0.684 | 0.666 |
| Sub 4 | 0.634 | 0.570 | 0.575 | 0.571 | 0.614 | 0.571 |
| pool | 0.657 | 0.603 | 0.640 | 0.603 | 0.651 | 0.603 |

**Table S10.** Results of observed vs. expected correctness based on source type

| Subjects | same source | different source | all |
|---|---|---|---|
| Sub 1 | – | – | – |
| Sub 2 | Pe < Po | Pe < Po | Pe < Po |
| Sub 3 | – | – | – |
| Sub 4 | Pe < Po | – | Pe < Po |
| pool | Pe < Po | Pe < Po | Pe < Po |

* Pe: expected performance; Po: observed performance.

**Table S11.** $\beta$ and $p$-value of observed vs. expected, based on equation (8)

| Subjects | same source | different source | all |
|---|---|---|---|
| Sub 1 | −0.023 ($p$ = 0.489) | −0.164 ($p$ = 0.247) | −0.067 ($p$ = 0.353) |
| Sub 2 | 0.389 ($p$ = 0.00019) | 0.669 ($p$ < 0.0001) | 0.481 ($p$ < 0.0001) |
| Sub 3 | 0.074 ($p$ = 0.374) | 0.069 ($p$ = 0.422) | 0.086 ($p$ = 0.279) |
| Sub 4 | 0.265 ($p$ = 0.0021) | 0.017 ($p$ = 0.5021) | 0.180 ($p$ = 0.0101) |
| pool | 0.233 ($p$ < 0.0001) | 0.157 ($p$ = 0.0254) | 0.209 ($p$ < 0.0001) |